\documentclass{article}
\title{Quantum mechanics and the continuum problem\\(with referee reports)}
\author{O. Yaremchuk\\email: yarem@online.ru}
\date{\today}
\begin{document}
\maketitle
\begin{abstract}

It is shown that Feynman's formulation of quantum mechanics can
be reproduced as a description of the set of intermediate
cardinality. Properties of the set follow directly from the
independence of the continuum hypothesis.

Six referee reports of Physical Review Letters, Europhysics Letters,
and Journal of Physics A are enclosed.

\end{abstract}

The concept of discrete space is always regarded as a unique
alternative to continuous space. Nevertheless, there is one
more possibility originated from the continuum problem: since
discrete space is a countable set, the set of intermediate
cardinality represents the golden mean between the opposing
concepts.

But although the continuum problem has been solved \cite{Cohen},
the status of the set of intermediate cardinality is still unclear.
The commonly held view is that, according to the independence of
the continuum hypothesis (CH), we can neither prove nor refute
existence of the intermediate set and, therefore, CH or its negation
must be taken as an additional axiom of standard Zermelo-Fraenkel set
theory (ZF). Thus we get two alternative set theories: with and without
the set of intermediate cardinality. The choice is a matter of
experimental verification analogously to the choice of true geometry
caused by the independence of fifth Euclid's postulate. In other words,
the independence of CH converts existence of the intermediate set into a
physical problem of determination of the set-theoretic structure of
space-time.

\medskip
Assume that the set of intermediate cardinality exists and consider
the maps of the intermediate set $I$ to the sets of real numbers $R$
and natural numbers $N$.

Let the map $I\to N$ decompose $I$ into
the countable set of mutually disjoint
infinite subsets: $\cup I_n=I$ ($n\in N$).
Let $I_n$ be called a unit set. All members of $I_n$
have the same countable coordinate $n$.

Consider the map $I\to R$.
By definition, continuum $R$ contains a subset $M$ equivalent to $I$,
i.e., there exists a bijection
\begin{equation}
f:I\to M\subset R.
\end{equation}
This bijection reduces to separation of the intermediate
subset $M$ from continuum. Since any separation procedure
is a proof of existence of the intermediate
set and, therefore, contradicts the independence
of the continuum hypothesis, we, in principle, do not
have any algorithm for assigning a real number to an
arbitrary point $s$ of the intermediate set. Hence, any
bijection can take the point only to a random (arbitrarily
chosen) real number. Thus we have the probability
$P(r)dr$ of finding the point $s\in I$ about $r$.

This does not mean that the intermediate set consist of random
numbers or that the members of the set are in any other sense
random. Each member of the set of intermediate cardinality equally
corresponds to all real numbers untill the mapping has performed
operationally. After the mapping, a concrete point gets random
real number as its coordinate (image) in continuum. It is clear that
we can get images of only finite number of points.

We see that the independence of CH may be understood as inseparability
of the subset of intermediate cardinality from continuum. Recall that
the intermediate set should be a subset of continuum by definition.

Thus the point of the intermediate set has two coordinates:
a definite natural number and a random real number:
\begin{equation}\label{s}
s:(n,r_{random}).
\end{equation}
Only the natural number coordinate gives reliable
information about relative positions of the points
of the set and, consequently, about size of an interval.
But the points of a unit set are indistinguishable. It is
clear that the probability $P(r)$ depends on the
natural number coordinate of the corresponding point.

For two real numbers $a$ and $b$ the probability
$P_{a\cup b}dr$ of finding $s$ in the union of the
neighborhoods $(dr)_a\cup (dr)_b$
\begin{equation}
P_{a\cup b}\,dr\ne [P(a)+P(b)]\,dr
\end{equation}
because $s$ corresponds to both (all) points at the
same time (the elemental events are not mutually exclusive).
In other words, the probability is inevitably non-additive.
In order to overcome this obstacle,
it is most natural to introduce a function $\psi(r)$
such that $P(r)={\cal P}[\psi(r)]$ and
$\psi_{a\cup b}=\psi(a)+\psi(b)$.
The idea is to compute the non-additive probability
from some additive object by a simple rule. It is quite
clear that this rule should be non-linear. Indeed,
\begin{equation}
P_{a\cup b}={\cal P}(\psi_{a\cup b})={\cal P}[\psi(a)+\psi(b)]\ne
{\cal P}[\psi(a)]+{\cal P}[\psi(b)],
\end{equation}
i.e., the dependence ${\cal P}[\psi (r)]$ is non-linear.
We may choose the dependence arbitrarily but the simplest
option is always preferable .
The simplest non-linear dependence is the square dependence:
\begin{equation}
{\cal P}[\psi(r)]=|\psi(r)|^2.
\end{equation} \label{born}

We shall not discuss uniqueness of the chosen options.
The aim of this paper is to show that quantum mechanics is,
at least, one of the simplest and most natural descriptions
of the set of intermediate cardinality.

The probability $P(r)$ is not
probability density because of its non-additivity.
This fact is very important.
Actually, the concept of probability should be modified,
since the additivity law is one of the axioms of the
conventional probability theory (the sample space should consist
of the mutually exclusive elemental events).
But we shall not alter the concept of probability because it is not
altered in quantum mechanics. 
This means that we shall regard $P(r)=|\psi(r)|^2$ as probability
density, i.e., we accept Born postulate. 

The function $\psi$, necessarily, depends on $n$: $\psi(r)\to\psi(n,r)$.
Since $n$ is accurate up to a constant (shift)
and the function $\psi$ is defined up to the 
factor $e^{i\mbox{const}}$, we have
\begin{equation}
\psi (n+\mbox{const},r)=e^{i\mbox{const}}\psi (n,r).
\end{equation}
Hence, the function $\psi$ is of the following form:
\begin{equation}
\psi (r,n)=A(r)e^{2\pi in}\label{debr}.
\end{equation}

Thus the point of the intermediate set corresponds to
the function Eq.(\ref{debr}) in continuum. We can specify
the point by the function $\psi(n,r)$ before the mapping
and by the random real number and the natural
number when the mapping has performed. In other words,
the function $\psi(n,r)$ may be regarded as the image
of $s$ in $R$ between mappings.

Consider probability $P(b,a)$ of finding the point $s$ at $b$
after finding it at $a$.
Let us use a continuous parameter $t$ for correlation
between continuous and countable coordinates of the point
$s$ (simultaneity) and in order to distinguish between
the different mappings (events ordering):
\begin{equation}
r(t_a),n(t_a)\to\psi(t)\to r(t_b),n(t_b),
\end{equation}\label{0-t}
where $t_a<t<t_b$ and $\psi(t)=\psi[n(t),r(t)]$.
For simplicity, we shall identify the parameter with
time without further discussion. Note that we cannot use the
direct dependence $n=n(r)$. Since $r=r(n)$ is a random number,
the inverse function is meaningless.

Assume that for each $t\in (t_a,t_b)$ there exists the image of
the point in continuum $R$. 

Partition interval $(t_a,t_b)$ into $k$ equal parts
$\varepsilon$:
\begin{eqnarray}
k\varepsilon =t_b-t_a,\nonumber\\
\varepsilon =t_i-t_{i-1},\nonumber\\
t_a=t_0,\,t_b=t_k,\\
a=r(t_a)=r_0,\, b=r(t_k)=r_k.\nonumber
\end{eqnarray}\label{partition}
The conditional probability of of finding the point $s$ at
$r(t_i)$ after $r(t_{i-1})$ is given by
\begin{equation}\label{cond}
P(r_{i-1},r_i)=\frac{P(r_i)}{P(r_{i-1})},
\end{equation}
i.e.,
\begin{equation}
P(r_{i-1},r_i)=\left|\frac{A_i}{A_{i-1}}e^{2\pi i\Delta n_i}\right|^2,
\end{equation}
where $\Delta n_i=|n(t_i)-n(t_{i-1})|$.

The probability of the sequence of the transitions 
\begin{equation}
r_0,\ldots ,r_i,\ldots r_k
\end{equation}
is given by
\begin{equation}
P(r_0,\ldots ,r_i,\ldots r_k)=\prod_{i=1}^k P(r_{i-1},r_i)=
\left|\frac{A_k}{A_0}\exp 2\pi i\sum_{i=1}^k\Delta n_i\right|^2.
\end{equation}
Then we get probability of the corresponding continuous sequence
of the transitions $r(t)$:
\begin{equation}\label{pathprob}
P[r(t)]=\lim_{\varepsilon\to 0}P(r_0,\ldots ,r_i,\ldots r_k)=\left|\frac{A_k}{A_0}e^{2\pi im}\right|^2,
\end{equation}
where
\begin{equation}
m=\lim_{\varepsilon\to 0}\sum_{i=1}^k\Delta n_i.
\end{equation}

Since at any time $t_a<t<t_b$ the point $s$ corresponds to all points
of $R$, it also corresponds to all continuous random sequences of
mappings $r(t)$ simultaneously, i.e., probability $P[r(t)]$ of finding
the point at any time ${t_a\leq t\leq t_b}$ on $r(t)$ is non-additive too.
Therefore, we introduce an additive functional $\phi[r(t)]$.
In the same way as above, we get
\begin{equation}
P[r(t)]=|\phi[r(t)]|^2.
\end{equation}

Taking into account Eq.(\ref{pathprob}), we can put
\begin{equation}\label{phi}
\phi[r(t)]=\frac{A_N}{A_0}e^{2\pi im}=\mbox{const}\,e^{2\pi im}.
\end{equation}
Thus we have
\begin{equation}\label{pathsum}
P(b,a) = |\!\!\sum_{all\,r(t)}\!\!\mbox{const}\,e^{2\pi im}|^2,
\end{equation}
i.e., the probability $P(a,b)$ of finding the point $s$ at $b$
after finding it at $a$ satisfies the conditions of Feynman's approach
(section 2-2 of \cite{Feynman}) for $S/\hbar=2\pi m$. 
Therefore,
\begin{equation}
P(b,a)=|K(b,a)|^2,
\end{equation}
where $K(a,b)$ is the path integral (2-25) of \cite{Feynman}:
\begin{equation}\label{pathint}
K(b,a)=\int_{a}^{b}\!e^{2\pi im}D r(t).
\end{equation}
Since Feynman does not essentially use in Chap.2 that $S/\hbar$ is just
action, the identification of $2\pi m$ and $S/\hbar$ may be postponed.

In section 2-3 of \cite{Feynman} Feynman explains how
the principle of least action follows from the dependence
\begin{equation}\label{sum}
P(b,a)= |\!\sum_{all\,r(t)}\!\!\mbox{const}\,e^{(i/\hbar)S[r(t)]}|^2.
\end{equation}
We can apply the same reasoning to Eq.(\ref{pathsum}) and,
for very large $m$, get ``the principle of least $m$''.
This also means that for large $m$ the point $s$ has a definite
stationary path and, consequently, a definite continuous coordinate.
In other words, the corresponding interval of the intermediate
set is sufficiently close to continuum (let the interval be
called macroscopic), i.e., cardinality of the intermediate set depends
on its size. Recall that we can measure the size of an interval of
the set only in the unit sets (some portions of points).

Since large $m$ may be considered as continuous
variable, we have
\begin{equation}\label{lim}
m=\lim\limits_{\varepsilon\to 0}\sum_{i=1}^k\Delta n_i=\int_{t_a}^{t_b}\!\!dn(t)=\min.
\end{equation}
The function $n(t)$ may be regarded as some function of
$r(t)$: $n(t)=\eta[r(t)]$. It is important that $r(t)$ is not
random in the case of large $m$. Therefore,
\begin{equation}\label{f}
\int_{t_a}^{t_b}\!\!dn(t)=\int_{t_a}^{t_b}\!\frac{d\eta}{dr}\,\dot{r}\,dt=\min,
\end{equation}
where $\frac{d\eta}{dr}\,\dot{r}$ is some function of $r$, $\dot{r}$,
and $t$. This is a formulation of the principle of least action
(note absence of higher time derivatives than $\dot{r}$), i.e.,
large $m$ can be identified with action.

Since the value of action depends on units of measurement, we need
a parameter $h$ depending on units only such that
\begin{equation}
hm=\int_{t_a}^{t_b}\!\! L(r,\dot{r},t)\,dt=S.
\end{equation}

Finally, we may substitute $S/\hbar$ for $2\pi m$ in Eq.(\ref{pathint})
and consider Feynman's formulation of quantum mechanics as a natural
description of the set of intermediate cardinality.

\medskip
Of course, the large intermediate interval is not exact continuum. The
principle of least action is an indication of intermediate cardinality of
the interval. Exact continuum does not have the principle as its inherent
property.

We see that if a point has sufficiently long countable path, then it has a
stable continuous coordinate. In other words, sufficiently long, in the
sense of the countable coordinate, interval of intermediate cardinality has
stable length, unlike unstable length of a small interval, i.e., length of
an intermediate interval becomes stabilized with increase in its cardinality.

Since length is establishment of euipotence $(r_1,r_2)\leftrightarrow R$,
if we reduce size of a large intermediate interval $(r_1,r_2)$, its length
becomes unstable and then collapses when cardinality of the interval
becomes sufficiently different from cardinality of continuum,
i.e., we have three basic kinds of the interval:

Macroscopic interval. This interval is large enough to be regarded as
continuous. It has stable non-zero length.

Microscopic interval. This interval may not be regarded as
continuos. In other words, it has no length, i.e., its continuos
image is exactly a point. 

Submicroscopic interval. It is an intermediate
kind of the interval with unstable random length. Submicroscopic
intervals make the region of quantum mechanics.

Thus the intermediate set is a set of variable infinite cardinality.
Taking into account that any infinite set should be equivalent to its proper
subset, we get that the set should have constant cardinality ranges, i.e.,
intermediate cardinality changes stepwise. Addition of only large enough
``portion'' of points changes cardinality of an intermediate subset to the
next level. It is reasonable to identify these portions with the unit sets.

Note that only sufficiently small intermediate interval manifests
explicit features of intermediate cardinality (the instability of length).
Such an interval may be called microscopic.
Existence of the ranges of constant cardinality makes possible equivalence
relations (symmetries) inside these ranges. This fact determines
applicability of symmetry groups which is important for microscopic subsets
because of absence of stable metrics and geometric structure. Outside a
subset of certain cardinality its internal symmetry should be broken.

From macroscopic point of view, there are two kinds
of points: the true points and the composite points. A composite
point (the microscopic interval) consist of an infinite number of
the true points. It is uniquely determined by the natural number of unit
sets. Cardinality of the proper microscopic interval may be regarded as
some qualitative property of the point (charge).
If the interval is destroyed (decay of the corresponding point),
this property vanishes and turns to the properties (cardinalities)
of the output intervals.

Thus the description of the proper microscopic intervals reduces to
the description of transmutation of expanded (non-local) but, at the
same time, point-like objects and their properties.

We see that the complete description of the set of intermediate
cardinality falls into a chain of three theories: classical mechanics,
quantum mechanics, and the description of the proper microscopic (point-like)
intervals whose properties are analogous to particle phenomenology
(note that we do not import any structure or law into the intermediate set).

At present, all the descriptions are imbedded in the continuous space of
classical mechanics. As a result, the dimensions and the directions
of the submicroscopic and microscopic descriptions are lost. 

The total number of space time dimensions of three 3D descriptions
is ten. The same number of dimensions appear in string theories.
But the extra dimensions of the intermediate set are essentially
microscopic and do not require compactification.

The directions of the submicroscopic and microscopic descriptions
are replaced with spin. Reliable separation of the descriptions
needs careful examination but it may be preliminarily stated that
integer spin is the direction of the microscopic description and
half integer spin is the direction of the submicroscopic one.
Since the submicroscopic interval is the (unstable) continuous
interval, its direction is associated with the direction of the
macroscopic continuous interval. Therefore, the submicroscopic
direction (half integer spin) is not a vector of full value but
only spinor.

Since microscopic intervals are substantially non-equivalent, the proper
microscopic description should in turn split into several ``asymmetric''
parts (different theories) with additional extra dimensions down to the
single unit set.

Fermions and bosons obey different statistics because they belong
to the different descriptions. The Pauli exclusion principle
is a condition for keeping inside the submicroscopic description
(in other words, this is just a condition of conservation of
submicroscopic cardinality): if two points at a submicroscopic
distance come close enough, in the sense of the countable coordinates,
they form the proper microscopic interval and go over to the proper
microscopic description. In this case, some macroscopic and
submicroscopic properties of the points of the interval may be lost.

Each microscopic scale should have analogous condition of conservation
of its cardinality, i.e., the law of conservation of some qualitative
property (charge). Violation of this law means conversion of initial
cardinality into cardinality of another proper microscopic scale.

The physical description of nature falls into a collection of different
theories steadily resisting  unification. The complete description of
the intermediate set exhibits the same tendency.
This is a consequence of the inherent structural nonuniformity of the set.
Thus instead of expected new fundamental principles, we get a new
fundamental object. The different fundamental theories are the legitimate
component parts of its complete description.

\subsection*{Supplement}
\noindent All the papers on this subject

(Interpretation of quantum mechanics and the structure
of physical space [1975],

Quantum phenomenology and the continuum problem,

Quantum mechanics and the continuum problem,

Quantum mechanics without interpretation,

The status of the set of intermediate cardinality)

\noindent were rejected by

Voprosy filosofii,

Kratkie so'obschenia po fizike,

Physical Review Letters,

Journal of Physics A,

Europhysics Letters,

Physics Letters A,

Foundation of Physics Letters,

The European Physical Journal C,

International Journal of Mathematics and Mathematical Sciences

Journal of Experimental and Theoretical Physics Letters

(Pis'ma v Zhurnal Eksperimental'noi i Teoreticheskoi Fiziki).

Available referee reports (Physical Review Letters, Europhysics Letters,
Journal of Physics A) are given below. Note that the e-print versions of
the papers differ from the more moderate versions sent to the journals.
My remarks I put into square brackets.

\subsection*{Physical Review Letters}

Second report of Referee A {\it[after resubmission.- O.Y.]}

Fri, 31 Mar 2000

Report on "Quantum mechanics and the continuum problem",

by O. Yaremchuk, LB7485.

   Although the paper still contains aspects and
assumptions that should be clarified, which is somehow
unavoidable for such an original proposal, I nevertheless
noticed the author's considerable effort in improving both
the content and the presentation. Overall I consider this
paper based on an underlying idea that, perhaps with some
modification, deserves some attention as it relates
several basic aspects in a rather new perspective. Thus,
considered also the preliminary nature of the
investigation, I recommend it for publication in PRL.

\medskip
Referee C

Referee's report:

   The referenced manuscript raises the question: Is
there a relationship between the continuum hypothesis and
quantum mechanics? The continuum hypothesis was originally
stated by George Cantor in the mid 19th Century {\it[in 1878 - O.Y.]}.
It hypothesizes the non-existence of sets
of intermediate cardinality between the cardinality of the
natural numbers and the cardinality of the real numbers
(the continuum). In 1963, the American mathematician Paul
Cohen established the independence of the continuum
hypothesis from the other axioms of Zermelo-Fraenkel set
theory. As it is independent, the negation of the
continuum hypothesis is consistent with Zermelo-Fraenkel
set theory. It is the resulting set of "intermediate"
cardinality that the author wishes to connect to quantum
mechanics.

   This is an intriguing idea. It raises the possibility
of two outcomes:

   *The intermediate set can be used to inform us of
quantum mechanics.

   *Quantum mechanics can be used to give us a novel
set theory.

   It seems that the former is the possibility that the
author tries to address. If he had done so, such a paper
would warrant publication in Physical Review. Addressing
the latter possibility would produce the paper on the
philosophy of mathematics (not a physics paper).

   It is my sad duty that the author does not reach his
desired goal; i.e., he does not show that the existence of
an intermediate set provides any insight (new or old) to
quantum mechanics. His paper fails on at least five
counts:

   1. 1. There is no discussion of the continuum
hypothesis nor Zermelo-Fraenkel set theory. I have
provided more in the first paragraph than this author
manages in nine pages.

   2. 2. He does not establish what the intermediate set
tells us about quantum mechanics; there is no insight
given about quantum mechanics.

   3. 3. His reasoning is, at best, non rigorous. {\it[The rigor
         of the paper quite conforms to that of Feynman's `Quantum
         mechanics and Path Integrals.'-O.Y.]}

   4. 4. He uses unexplained novel methods.

   5. 5. He gets some things plain wrong.

{\it[The way of numbering of above items is original.-O.Y.]}

   I now provide examples of each of the five failures
listed above:

   As to the first count: At a minimum the author should
have sited references for the relevant ideas in set
theory. The two references he cites are both physics
texts. {\it[I added the well-known Cohen's book.- O.Y.]}

   As to counts two and three: The closest that the author
comes to providing insight is the following passage on
page two

   "Then some subset cannot be separated from continuum
if each point of the subset does not have its own peculiar
properties but only combines properties of the members of
the countable set and continuum.

At first sight, this seems to be meaningless.
But the content of the requirement coincides with the
content of wave-particle duality: quantum particle
combines properties of a wave (continuum) and a
point-like particle (the countable set)."

   This statement comes out of the blue. Instead of
telling us something about quantum mechanics he seems to
be saying: "Quantum mechanics is weird; the intermediate
set is weird; ergo, one must be related to the other." If
anything, the author is trying to use quantum mechanics to
provide insight to sets of intermediate cardinality,
rather than the other way round. {\it[I think that the analogy
between wave-particle duality and intermediate cardinality is
obvious.-O.Y.]}

   Another illustration of the lack of insight and rigor
comes on page three in the following passage:

   ``The simplest non-linear dependence is a square
dependence:
\begin{equation}
{\cal P}[\psi(r)]=|\psi(r)|^2."
\end{equation}

   What is his justification for saying this?  I submit
it is precisely because it is the mathematical choice that
gives us quantum mechanics. Here I see no insight being
provided in either direction; there is only a bald
assertion that leads to the author desired result.

{\it[Unfortunately, referee C does not inform us of his discovery of
the simpler non-linear dependence. In fact, I did not make even this choice.
This is a choice of the founders of quantum mechanics justified by
experiment. I only show that their (perhaps arbitrary, perhaps uniquely
determined) sequence of actions is suitable and quite natural in order to
reproduce quantum mechanics as a description of the intermediate set.
I have to prove that quantum mechanics describes the set of intermediate
cardinality but it is not necessary to insist that this is a unique or the
best way to describe the set.-O.Y.]}

   As to count four: A prime example is found on page
three:

   ``Hence, any bijection can take a point of the
intermediate set only to a random real number. If we do
not have preferable real numbers, then we have the
equiprobable mapping. This already conforms to the quantum
free particle.''

   I have no idea what is meant by any mapping (much less
a bijection) to a ``random real number.'' I can describe
the class of random number by algorithmic methods but I
have no idea how to set up a mapping into this class.

As a mapping must involve a definite rule, I see no way
to get what the author desires here. For him to then
discuss an image under this mapping as having the form
(natural number, random real) simply leaves me speechless.
Maybe he has something in mind here; if so, he needs to
explain it carefully and rigorously.

{\it[Referee C seems to look for some subtle sense.
In fact, the meaning is rather primitive.
If one takes tree balls out of the urn which contains ten balls (and if
one is interested only in the number of members in the separated subset
which is equivalent to absence of any selection rule), one establishes a
bijection:

${(1,2,3)\to (\mbox{three random balls})\subset(\mbox{ten balls})}$.

If our urn contains ten black balls and ten white balls, we can take three
white balls, that is, we get random balls with the non-random property.
Analogously, we can control, theoretically, the natural number
coordinate and cannot the real number one.-O.Y.]}

   As to count five: A prime illustration occurs on
page eight:

   ``The total number of space time dimensions of
three 3D descriptions is ten. The same number of
dimensions appear in string theories.''

   The number of spatial (not space-time) dimensions is
ten. This is the most intelligible statement in the last
two pages and he gets it wrong.

{\it[This is a quotation of John H. Schwarz's ``Introduction to
Superstring Theory,'' hep-ex/0008017 (p.7): ``...that there are five
distinct consistent string theories, and that each of them
requires spacetime supersymmetry in the ten dimensions (nine
spatial dimensions plus time).'' Such a mistake of referee and editor
(George Basbas) of such a respectable journal seems to me
inexplicable.-O.Y.]}

   This paper in no way qualifies for publication
anywhere. I can not imagine any possible revision making
this a publishable piece of work. This is a true pity, as
I think the author has the kernel of an interesting idea.

\subsection*{Europhysics Letters}

G11661

"QUANTUM MECHANICS AND THE CONTINUUM PROBLEM"

\medskip
REPORT A

The author assumes the existence of a set
of intermediate cardinality, between continuous
and discrete spaces.
Then he derives a connection between this set
and many features of quantum mechanics.
In particular, he shows how a wave function
can be introduced, and how it is related to
a probability density.
Moreover, some properties of the functional
integral formulation of quantum mechanics
are given the interpretation in this
particular framework.
The essential content of the paper is highly
speculative, and it is not clear how the
proposed framework could be useful in some
concrete problem.
However, the paper contains new ideas, quite
stimulating. It could be interesting to let
them be known to the scientific community.
Therefore, I am inclined to recommend publication
of this paper on Europhysics Letters, on
the basis of its interesting and stimulating
original conceptual content.
On the other hand, due to the speculative character,
this is a case where the decision involves
directly the editorial policy of the Journal.

{\it[It is interesting that discreteness and continuity are not
considered as speculative in contrast to the mean concept.-O.Y.]}

\medskip
REPORT B

REPORT ON THE PAPER "QUANTUM MECHANICS AND THE CONTINUUM PROBLEM"

\medskip
The paper proposes an approach to quantum mechanics based on the
independence of the continuum hypothesis.
The independence of the continuum hypothesis means that
the statement "there does not exist a subset of the reals
whose cardinality is intermediate between that of the
integers and that of the reals" is not deducible from
the axioms of set theory. One of the proofs consists
in constructing a model of real numbers in which
sets of intermediate cardinality exist (see for example
Yu. Manin "A course in mathematical logic" Ch. III).
My impression is that the author of the paper is not sufficiently
familiar with the continuum problem.

{\it[Not `one of the proofs' but the only proof of the independence
of CH consist in constructing models of ZF (not real numbers)
with (Cohen) and without (Goedel) the set of intermediate cardinality.
Yu. Manin only explains Cohen's idea by model of real numbers
(the language of real numbers is simpler than that of ZF).
Unlike Goedel model, which is really `smallest set theory' consisting only
of constructible sets (constructible universe), the model with the
`intermediate set' is an unnatural extension of set theory.
Of course, the real set of intermediate cardinality was not constructed
contrary to the false conclusion of the referee. The incomprehension
stated below also follows from this misunderstanding.-O.Y.]}

The crucial  sentence after eq. (1)

'Since any separation rule is a proof of existence of the
intermediate set and , therefore, contradicts the independence
of the continuum hypothesis, we, in principle do not
have a rule for assigning  a definite real number
to an arbitrary point of the intermediate set.'

is wrong for the first part and meaningless for the second.
All subsequent developments look to me confused and arbitrary
and I am  not able to  see the pretended connection between quantum
mechanics and the continuum problem.

The paper does not meet acceptable standards of
scientific communication and I recommend rejection. 

\subsection*{Journal of Physics A: Mathematical and General}

FIRST REFEREE REPORT

A/113435/PAP

Quantum mechanics without interpretation

O Yaremchuk

\medskip
1\quad
As the configuration space for quantum mechanics, the author proposes the
existence of a unique set of intermediate cardinality between the set of
integers and the continuous space. Because of the independent of continuum
hypothesis (CH), the negation of CH can be taken as an axiom. {\it[The set
of intermediate cardinality is not a configuration space.-O.Y.]}

2\quad
Because of independent of CH, the intermediate set can not be described by
some formula. Therefore the author argues that the set as a subset of
continuous space must have combined properties of the members of countable
set and continuum, which coincide the content of wave particle duality in
quantum mechanics.

3\quad
The author gives an argument with which the properties of the
intermediate set can be described by Feynman's path integral method.

4\quad
However the referee thinks that the configuration space of quantum mechanics
should be constructive and can be described mathematical formula.
Philosophically the referee can not accept the author's approach.

{\it[Note that the reason for rejection of the paper is
philosophical.-O.Y.]}

\medskip
SECOND REFEREE REPORT

Quantum mechanics without interpretation

O Yaremchuk

A/113435/PAP

\medskip
The author suggests that the quantum wave-particle duality
arises when a system has a phase space of cardinality
intermediate between $aleph_0$ (the cardinality of the integers) and $C$
(the cardinality of the real numbers). He calls such sets
``intermediate sets.'' {\it[This is very peculiar understanding which I
cannot confirm. The intermediate set is not a phase space. Duality means
that the intermediate set combines properties of the continuous and countable
sets and have not any exclusive property, therefore, any description (and its
phase space) of any intermediate set contains only continuous and discrete
(quantized) quantities.
The non-mathematical concept of wave-particle duality was introduced into
quantum theory because of absence of adequate mathematical notion.
I try to explain this duality in terms of intermediate cardinality in order
to establish the complete coincidence between the the intermediate set
and the mathematical object described by quantum mechanics.-O.Y.]}

He implicitly proposes that Bohr undecidability is a consequence of
Goedel undecidability. This proposal has often been made, but has rarely
resulted in printable papers. {\it[One reference would be enough.-O.Y.]}
Yaremchuk carries it further than anyone else has, so far as I know.
{\it[Anonymous referee makes anonymous references.
No one else ever stated that quantum mechanics described the set of
intermediate cardinality. Moreover, my first attempt to discuss relationship
between quantum mechanics and the continuum problem with mathematicians
in sci.math (deja news) caused the following response:
extremely silly...keep your idiocy off my mail box...-O.Y.]}

His title is exactly the opposite of his paper. What he proposes is
exactly what those who speak of ``interpretations'' of quantum mechanics
mean. {\it[Interpretation is necessary in order to explain what is really
described by quantum mechanics. I start with the definite mathematical
object which is sufficiently primitive to be regarded as real (in contrast
to Hilbert space or operator algebra).-O.Y.]}
Undoubtedly the title responds to a recent paper of the same title by
Peres et al. in Physics Today. {\it[I tried to find the paper `of the same
title,' that is, ``Quantum mechanics without interpretation.'' I found
only ``Quantum mechanics needs no `interpretation' '' by Fuchs and Peres.
This paper presents very radical (`empty') interpretation.-O.Y.]}
But the point of Peres is that works like this paper are not necessary in
order to make quantum theory satisfactory. I agree with Peres. 

I do not share the paper's goal or philosophy.
In my opinion the author is misled by a superficial resemblance between the
two kinds of undecidability. Goedel undecidability is a limit to the power
of the postulational method, which generates fewer theorems than
decidability requires. Bohr undecidability is a revision of the
operator algebra. The author is simply clinging the old postulates ---
selective operation must commute, complete description must exist ---
in the face of new experience. The paper is of the same general class
as papers showing that in spite of special relativity there might still
be an absolute time, or paper showing that in spite of general relativity
there might still be absolute acceleration. Such theories generally
accomplish their goals by introducing quantities that allegedly ``exist
but are unobservable.'' Here the unobservable is the point of the
intermediate set. {\it[I did not introduce the concepts of ``observable''
or ``unobservable.'' This blame my be rather laid on the founders of
quantum mechanics. The ``observable'' principle of least action and quite
``observable'' quantum-mechanical formalism, which describes the behavior
of the point, are the indications of intermediate cardinality of space
to which the point belongs.-O.Y.]}

Certainly the present paper does not succeed in deriving the basic
principles of quantum kinematics from the hypothesis of the intermediate
set. The key formula (5) still has to be postulated, as it is in most
presentations of quantum theory. In quantum theory (5) actually does not
need to be postulated separately but follows from assumed simplicity of the
operator algebra. {\it[The problems and peculiarities of quantum-mechanical
way of description, which I must reproduce, and the problems with the
intermediate set as a new object are regarded as weak points of my paper.
In the new context of intermediate cardinality, it becomes obvious that
postulate is not the best way to introduce the probability density.
But this is not my idea. The problem needs careful analysis.-O.Y.]}

The author identifies physical processes like measurements with
mathematical mappings like deductions. No evidence is offered for a
structural similarity between measurements and deductions.
{\it[Mathematical mapping is a general concept relating to any
special case. Since any additional condition is not required,
our reasoning is valid for any realization of mapping including
measurement and natural phenomena without observers.-O.Y.]}

The author proceed as if the wave-function psi were a property of the
system like coordinate or an electric field. In quantum theory it is not,
but describes a process by which a system can be prepared or registered.

The identification (in the text between (3) and (4)) of the union of
neighbourhoods with the addition of wave-functions psi is a simple blunder.
It is consistent with the author's own understanding that the wave-function
is only determined up to phase. {\it[Note that no one of the other
referees has paid attention to this ``blunder.''-O.Y.]}

The infinities introduced by this hypothesis make it unlikely that a
convergent physical theory will result. It is possible that all the
thermal infinities that quantum theory avoids will plague this theory.

I recommend that the paper be declined as not sufficiently correct.
I would encourage the author to pursue his investigation of the
physics of the intermediate sets until he produces a more rigorous study,
with more attention to the roles and interpretations of the elements
of quantum theory. His hypothesis is intriguing and will continue to
arise until a definitive work settles it one way or the other. He has
gone further with it than anyone else so far.

A minor points: The writing needs improvement. The articles ``a'' and
``the'' seem to occur rather randomly and impede the reading. The
author is carried away by enthusiasm in the later pages of the paper,
presenting his hopes as if they were facts. They should either be
supported by proofs or presented as conjectures.
{\it[The main difficulty is that intermediate cardinality seems intuitively
unclear pure mathematical concept. Therefore, in the latter pages, I try to
show that the intermediate set gives convincing informal picture and can
generate the same peculiar concepts that have been introduced into
distinct quantum theories for consistency with experiments (wave-particle,
spin, two kinds of spin, exclusion principle, charges of particles, strings,
extra dimensions, confinement, etc) and clarify physical meanings of all
the concepts the only explanation of which is `absence of classical analog'.
Detailed coincidence of informal pictures cannot be accidental, i.e., all the
quantum phenomena unambiguously point to the set of intermediate cardinality.-O.Y.]}

\end{document}